\documentstyle[12pt,psfig]{article}

\begin{document}
\thispagestyle{empty}

\mbox{}
\vspace{1cm}
\begin{center}
{\bf Modelling coevolution in multispecies communities} \\

\vspace{0.5cm}

{\it Guido Caldarelli \footnote[0]{Present Address TCM, Cavendish Laboratory
Madingley Road, Cambridge CB3 0HE UK}}$^{,1,2}$, {\it Paul G. Higgs}$^2$ {\it and \
Alan J. McKane}$^1$ \\
\bigskip
$^1$ Department of Theoretical Physics, University of Manchester, \\
Manchester M13 9PL, UK \\

\medskip

$^2$School of Biological Sciences, University of Manchester, \\
Manchester M13 9PT, UK
\end{center}

\vspace{2cm}

\begin{abstract}

\bigskip

We introduce the Webworld model, which links together the ecological 
modelling of food web structure with the evolutionary modelling of
speciation and extinction events. The model describes 
dynamics of ecological communities on an evolutionary timescale.
Species are defined as sets of characteristic features, and these features 
are used to determine interaction scores between species. 
A simple rule is used to transfer 
resources from the external environment through the food web to each of the 
species, and to determine mean population sizes. 
A time step in the model represents a speciation event. 
A new species is added with features similar to those of one of the existing 
species and a new food web structure is then calculated. The new species may
(i) add stably to the web, (ii) become extinct immediately because it is poorly
adapted, or (iii) cause one or more other species to become extinct due to
competition for resources.

We measure various properties of the model webs and compare these with
data on real food webs. These properties include the  
proportions of basal, intermediate and top species, the 
number of links per species and the number of trophic levels.
We also study the evolutionary dynamics of the model ecosystem by following the
fluctuations in the total number of species in the web. Extinction 
avalanches occur when novel organisms arise which are significantly better 
adapted than existing ones.
We discuss these results in relation to the observed extinction events 
in the fossil record, 
and to the theory of self-organized criticality.	
\end{abstract}

\newpage
\pagenumbering{arabic}

\section{Introduction}

Our aim in this paper is to introduce a model for the evolution of many 
interacting species in an ecosystem. A great deal of information has been 
assembled on the 
properties of food webs for naturally occurring ecosystems (Cohen, 1989;
Martinez, 1991; Polis, 1991; Hall \& Raffaelli, 1991; Goldwasser \& Roughgarden,
1993), and a variety of models describing the 
structure of food webs have been studied (Pimm, 1982; Briand \& Cohen, 1984;
Cohen et al., 1990; Cohen, 1990; Pimm et al., 1991, Morin \& Lawler, 1995). 
Most of these models
concentrate on static properties of the webs, such as the proportions of 
species at each of the trophic levels, and the lengths of food chains. There 
have also been models for the assembly of food webs by gradual addition of 
species (Luh \& Pimm, 1993; Morton \& Law, 1997). 
One essential property of real food webs is
that they have evolved. Species diversity will accumulate over time as new 
species arise due to speciation. Species may also become extinct due to 
competition from better adapted species. Here we study a model which has the 
evolutionary dynamics of speciation and extinction built into it. We will 
analyse the properties of the 
food webs which arise in the model and compare these with ecological data.

The variation in the diversity of species seen in the fossil record is well
documented (Sepkoski, 1993). Attention has focused on the large-scale
extinction events in which very large fractions of the existing species 
apparently
become extinct almost simultaneously on a geological timescale. Some of these
large scale extinctions may be caused by catastrophic external events, or by
major climate changes. 
However it has been shown (Sneppen et al., 1995; Sol\'e et al., 1997)
that both the
distribution of sizes of extinction events and the distribution of lifetimes
of species in the fossil record are very broad and approximate to power-laws. 
It has
been argued that the large fluctuations in species diversity which are seen may
result from the internal dynamics of the ecosystem. A newly-evolved species may
cause extinction of its immediate competitors. The extinction of one species 
may lead to the extinction of other species which prey upon it, and these 
extinctions may lead to further extinctions, etc. Hence there is the 
possibility of avalanches of
extinction events spreading through an ecosystem. A link has been made between
macroevolutionary dynamics and the theory of self-organized criticality (SOC).
This theory was originally formulated in the 
language of sandpiles (Bak et al., 1988)
and has since been applied to many other dynamical systems including 
earthquakes and forest fires. In the sandpile model one grain of sand is added 
at a time onto a pile of sand. This may either add stably to the pile, or may 
cause an avalanche of one or more grains to tumble from 
the pile. The evolutionary analogy is clear: evolution leads to new species 
being added slowly to an ecosystem, and these may sometimes add stably to the 
system or may sometimes cause extinction events.

There is now a huge variety of evolutionary models inspired by the idea of SOC
(Bak \& Sneppen, 1993; Paczuski et al., 1996; Kramer et al., 1996;
Sol\'e \& Bascompte, 1996; Sol\'e et al., 1996; Roberts \& Newman, 1996). 
These models are deliberately extremely simple, and they focus on the dynamical 
properties of extinction avalanches and species lifetimes. They do not tell us 
anything about ecosystem structure, because the couplings between species are 
introduced either at random or in an ad-hoc manner (e.g. species are arranged on
a line and neighbours interact). In our view it is essential to have a realistic
model of ecosystem structure in order to study the way extinction events will
propagate through a food web. By linking the ecological modelling of food web 
structure with the evolutionary modelling of speciation and extinction events 
we believe that the Webworld model described in this paper
makes a significant advance in both fields.

At this stage it is useful to introduce some notation used in the description
of food webs. If one species
preys on another then they are said to be {\it linked}. {\it Basal species} 
are those with predators 
but with no prey and {\it top species} are those with prey but with no 
predators. {\it Intermediate species} have both predators and prey. We will
refer to the percentages of basal, intermediate and top species as $B$, $I$, and 
$T$.
Two species are said to belong to the same {\it trophic species}
if they share the same set of prey and the same set of
predators (see Figure 1). It is customary to group the original species in a
web into trophic species and to analyse the properties of the trophic species 
web. This goes some way toward standardizing webs obtained by different 
researchers using different criteria for inclusion of species and 
different degrees of resolution. The statistics presented in this paper will
be for trophic species webs. One food web statistic which we quote for the model
webs is the ratio of prey to predator species, defined as $(B+I)/(T+I)$.
Another characteristic 
property of webs is the average number of links per species, defined simply 
as the total number of links divided by the total number of species. It is also
useful to classify the links according to the type of species which they 
connect, i.e. we can measure the
proportion of links between top and basal ($TB$), top and intermediate
($TI$), intermediate and intermediate ($II$) and intermediate and basal ($IB$) 
species. 

 Ecosystems are reliant on the input 
of resources from the external, non-living environment. Within the model the 
environment is treated explicitly as a node in the food web. Species which are
linked to the environment node are primary producers, i.e. they may survive in
absence of any other species. It follows that basal species (those having no
prey) must be primary producers, otherwise they would have no means of
survival. However not all primary producers are basal species, since a primary
producer may also be a predator of other species.

We will use the following definition of {\it trophic levels}. All 
species linked to the environment node (i.e. primary producers) are
defined as level 1 species.
All species which have at least one level 1 prey are defined as level 2
species. All species with at least one level 2 prey are defined as level 3, and
so on. Hence, the level of a species is the length of the shortest food chain 
from the
external environment to that species. We could also have chosen to characterize
species by the length of the longest chain linking them to the environment,
 or the average length of all the possible chains, however these quantities are 
time consuming to calculate for large webs, and also special rules are
required to deal with the possibility of loops in the web. The definition we use
is rapid to calculate, and is unambiguous, even when loops are present in the
web. Also the majority of resources obtained by
a species are likely to come via the shortest route, since transfer of 
resources 
at each link of the food chain is relatively inefficient. Therefore it seems that
the definition of trophic level which we use is meaningful from an
ecological point of view.

\section{Definition of the Webworld Model}
\noindent{\bf Features and Species} 

\medskip

A species is defined in terms of a set of $L$ characteristic features.
Such features may be either morphological or behavioral characteristics
possessed by all the individuals of that species, e.g. sharp teeth, binocular
vision, webbed feet, being nocturnal, forming social groups etc. The $L$ 
features for each species are picked from a total number of $K$ possible 
features. In the simulations presented here $L = 10$ and $K = 500$. The species
are labelled by integers $n, n' = 1, 2,...$.

The predator-prey relationships between species are determined by 
the features possessed by those species. 
The matrix, $m$, is a $K\times K$ matrix of scores representing the usefulness 
of any feature
$i$ of one species for predation against any feature $j$ of another species. We
suppose that $m$ is anti-symmetric, and that the elements are assigned 
randomly, such that $m_{i,j}$ is a random Gaussian variable with mean zero and
variance 1 if $i > j$, $m_{j,i} = -m_{i,j}$, and the diagonal elements
$m_{i,i}$ are zero. 
This matrix of the scores is chosen at the beginning of a simulation run and 
does not
change during the run. The score $S_{n,n'}$ of one species $n$ against another
species $n'$ is then defined as
\begin{equation}
\label{species}
	     S_{n,n'}=max\{0,\frac 1 {L} \sum_{i \in n} 
                           \sum_{j \in n'} m_{i,j} \}
\end{equation}
where $i$ runs over all the features of species $n$ and $j$ runs over
all the features of species $n'$. The $max$ operator ensures that all scores 
are positive or zero. A positive score indicates that $n$ is adapted to be a 
predator of $n'$, and a zero score indicates that there is no interaction. 
The factor of $1/L$ has been chosen so that the scores have a root mean 
square value of order 1. 

The external environment is represented as an additional species $0$
which is assigned a set of $L$ features randomly at the beginning of a run
and which does not change. If $S_{n,0}>0$ the
species $n$ may be a primary producer, provided it is not out-competed
(see below).

\medskip

\noindent {\bf Transfer of Resources}

\medskip

A total amount, $R$, of external resources is distributed amongst the primary
producers, as a function of the scores $S_{n,0}$, according to the rules
for competition described below. These resources determine
the population size of the primary producers. The population of species $n$ is
denoted $N(n)$. For simplicity, we measure resources and population size in the
same units, so that $N(n)$ is equal to the amount of resources obtained by 
species $n$. Predator species
obtain resources from their prey. We assume that a fraction $\lambda N(n)$ of
the resources of a given species $n$ is available to be passed on to predators 
of $n$. Thus $\lambda$ is a parameter of the model which determines the 
relative sizes of predator and prey populations.

\medskip

\noindent{\bf Competition for Resources} 

\medskip

The resources obtained from species $n$ are distributed between
the predators of $n$ in the following way. (The same rules are used for
the case $n = 0$, where the
resources are the external resources $R$, and the `predators' are the
primary producers). The better adapted
the predator, the more resources it gets. We therefore define the {\it main
predator} of $n$ as the one with the best score against $n$:
\begin{equation}
\label{main}
  S^{M}_{n}=max\{S_{n',n}\} \ \ {\rm where \ } n' \
{\rm is \ a \ predator \ of\ } n
\end{equation}
The predators of $n$ will obtain a share of the available resources
proportional to the quantity:
\begin{equation}
\label{compet}
    F_{n',n}=max\left\{ 0,\left(1-\frac{S^{M}_{n}-S_{n',n}} 
	     {\delta}\right) \right \}
\end{equation}
where $\delta$ is a parameter of the model which determines the strength of
competition. The smaller $\delta$ the stronger the competition. We also
define $F_{n',n}$ to be zero if $S_{n',n} = 0$, so that only species for
which $S_{n',n}>S^{M}_{n} - \delta$, and $S_{n',n}>0$ successfully obtain 
any resources from $n$. The actual fraction of resources obtained is found by
normalizing the $F_{n',n}$:
\begin{equation}
\label{gamma}
\gamma_{n',n} = \frac{F_{n',n}} {\sum_{m} F_{m,n}}
\end{equation}

\medskip

\noindent{\bf Population Sizes}

\medskip

We will calculate the population sizes of the species
by iteration of a set of equations representing transfer of resources
between the species. Each iteration represents a small time period of order one
generation time. If $N(n,t)$ is the population of species $n$ at iteration $t$
then we may write
\begin{eqnarray}
\label{step1}
N(n, t+1) & = & \gamma_{n,0}R + \sum_{n'} \gamma_{n,n'}\lambda N(n', t) 
+ \gamma_{n,n}\lambda N(n, t).
\end{eqnarray}
The first term represents external resources obtained by $n$, which may be zero
if $\gamma_{n,0}=0$. The second term represents resources obtained from prey 
($n'$ runs over all the prey of $n$). The third term represents resources lost
to predators. Here we have defined
\begin{equation}
\label{extgamma}
\gamma_{n,n} = \left\{ \begin{array}{ll} 
-1, & \mbox{\ if $n$ has at least one predator species} \\ 
\ \ 0, & \mbox{\ if $n$ has no predators}
\end{array} \right.
\end{equation}
After several iterations the population sizes will converge to stationary
values. We suppose that the stationary population sizes 
determined by this method represent
mean population sizes over many generations.  
The iteration procedure can be viewed as an algorithm which determines 
the solution of the 
following set of linear equations for the stationary values:
\begin{equation}
\label{steady}
    N(n) = \sum_{n'} \gamma_{n,n'} \lambda N(n').
\end{equation}
Here $n'$ runs over all the species, including $n$ and 0, and for convenience
we have defined $N(0)=R/\lambda$.

There have been many studies of population dynamics at short timescales using
either difference equations like (\ref{step1}), or differential equations
like the Lotka-Volterra equations, or generalizations of them
(Vida et al., 1990; Berryman et al., 1995; Arditi \& Michalski, 1996; 
Zheng et al., 1997). Many of these equations have interesting
dynamical behaviour, such as periodic solutions or chaos which may be
relevant to real ecological population dynamics. In fact we are not interested 
in the population dynamics at the
scale of a few generations. We are only interested in evolutionary time.
We have deliberately
chosen difference equations which are as simple as possible, and which 
have only one stationary state. These equations
converge rapidly, and always reach the same stationary state irrespective of
the initial conditions. It is a property of our equations that the amount of
resources passed from a prey to a predator species is proportional to the prey
population size only. This situation is usually called donor control (Zheng et
al., 1997). We will be interested in the way the population
sizes change as the ecosystem evolves. 
Before describing the long timescale evolutionary
dynamics of our model it is useful to consider the following simple example
which illustrates competition between species and transfer of resources
between levels. 

Suppose there are two primary producers, 1 and 2, having scores $S_{1,0}=1.0$
and $S_{2,0}=0.95$. In addition, species 3 is a predator of species 1, i.e.
$S_{3,1}>0$. Let the competition parameter $\delta$ be 0.1. From the
above rules $\gamma_{1,0}=2/3$ and $\gamma_{2,0}=1/3$, and since 3 is the sole
predator of 1, $\gamma_{3,1}=1$. The iterative equations are therefore
\begin{eqnarray}
\label{example}
N(1,t+1) & = & 2R/3 - \lambda N(1,t) \nonumber \\
N(2,t+1) & = & R/3 \\
N(3,t+1) & = & \lambda N(1,t) \nonumber
\end{eqnarray}
and the stationary states are $N(1)=2R/3(1+\lambda)$, $N(2)=R/3$, and 
$N(3)=2\lambda R/3(1+\lambda)$. Notice that the sum of the populations is
equal to $R$. It is a property of the model that the sum of the populations at
the stationary state is always equal to the amount of external resources put in.

Given the lists of features representing any set of species it is possible to 
calculate the scores $S_{n,n'}$ and hence the steady state population sizes. We 
suppose that there is a minimum population size necessary for survival of
the species, and we set this limit to 1. Any species for which $N(n)<1$ becomes
extinct and is deleted from the list of species. The resulting ecosystem is then
stable, and properties of the web can be measured. 

For the purposes of
defining the food web structure, a link is assumed to be present between $n$ and
$n'$ if $n$ successfully obtains resources from $n'$, i.e. if $\gamma_{n,n'}>0$.
Cannibalism has been excluded from the model ($S_{n,n}$ is defined to be zero 
for all $n$). Also, 
since the $m_{ij}$ matrix is antisymmetric, then for any pair of species it is 
impossible for both $S_{n,n'}$ and $S_{n',n}$ to be non zero, i.e. there are no
reciprocal pairs of links. Loops of three or more species can occur in the model,
however.

\medskip

\noindent{\bf Evolutionary Dynamics}

\medskip

Evolutionary time in the model proceeds in timesteps such that one speciation
event occurs in every timestep. At each step an existing species is chosen 
to undergo speciation with a probability proportional to its 
population size. A new species is created by copying the list of features
of the parent species, randomly picking one these 
features, and replacing it by another randomly chosen 
feature from the complete list of possible features. The new species thus
differs by only one feature from the old one. The stationary population sizes 
of all 
the species are now recalculated, taking account of the presence of the new
species. There are several possible outcomes: (i) the new species may add to 
the web in a stable fashion, so that the total number of species increases by
one; (ii) the new species may be poorly adapted, and may become extinct
immediately; (iii) the new species may survive, and cause one or more other
species to become extinct. A new species will often be in direct competition 
with the parent species, therefore a special case of outcome (iii) is that the 
new species simply replaces its parent in the ecosystem.

Having eliminated any species which become extinct due to the new addition,
the web is once again in a stable state. This is the end of one complete
timestep. This process is repeated many times and the properties of the webs are
recorded at the end of each step and subsequently averaged. 
When choosing the new feature for each new species a restriction is made that
no feature may appear more than once in the list of features possessed by any 
one species. Also, when the new feature list is created, we check that it is 
not identical to the feature list of any other species, and that it is not 
simply a permutation of any other list. This prevents the occurrence of 
multiple copies of identical species.

The only remaining thing to be specified is the initial state of the web. One
possibility is to consider an `origin of life' scenario, where the web begins
with exactly one primary producer species. Another possibility is to begin 
with a small number of randomly chosen species. We carried out runs where the 
initial state was composed of 1, 10 and 20 different species, and found very 
little difference between these. All the results presented here were started 
with a set of 10 random species. Since the features of these species are 
chosen at random, the species are not well adapted to survive together, hence 
there is a tendency for several extinctions to occur immediately on the first 
timestep.

\section{Properties of the Model}

In this section we discuss some of the important properties of the model which
are relevant to understanding the numerical results in the next section. 
The model contains three parameters, $R$, $\lambda$ and $\delta$, and these 
will affect the shapes of the food webs in various ways. 
	
We would intuitively expect that increasing the amount of resources available
should increase the diversity of the ecosystem. In the model the sum of the 
population sizes is equal to $R$, and since every species must have a population
of at least one, there can never be more than $R$ species. In fact the mean
number of species observed is always much less than $R$ because resources are
not distributed evenly between species. Species at lower levels have much 
larger populations. The ratio of population sizes between predator and prey is
controlled by $\lambda$. Together $R$ and $\lambda$ determine the maximum 
possible number of trophic levels in the system.

Consider a web consisting of a single food chain of $k$ species. The stationary
population sizes obtained from the model for each level $j$ except the
top level are $N(j)=\lambda^{j-1}R/(1+\lambda)^{j}$, and the population of the 
top level is $N(k)=\lambda^{k-1}R/(1+\lambda)^{k-1}$. A chain of length $k$ can
only be supported if $N(k)>1$. This gives the condition
\begin{equation}
\label{maxlevel}
k < 1 + \frac{\log{(R)}} {\log{((1+\lambda)/\lambda)}}
\end{equation}
and, since $k$ is an integer, the maximum level is actually the largest integer
below this limit.
Thus the maximum food chain length only increases logarithmically with $R$. 
Although this limit has been calculated assuming the web is a single chain, 
we believe that this is a strict limit to the number of levels for any possible 
web generated by the model. This is because adding extra links and more 
species to the web always decreases the fraction of resources reaching the top 
level in comparison to the single chain calculation. 

The number of levels observed in the model food webs is strongly sensitive to
$\lambda$. We have chosen to set $\lambda = 0.1$ in all the simulations 
reported here. This gives realistic numbers of trophic levels in the model
food webs and is also consistent with measured values of
predator prey population ratios, and estimates of the ecological efficiency
(Pimm, 1982).

The value of the parameter $\delta$ affects the
properties of the web considerably. In order to choose a sensible value of
$\delta$ we note that the scores $S_{n,n'}$ are all of order 1, and that when a
single feature is changed, the score changes by an amount of order $1/L$. 
In other words, species competing for the same resources should have scores
which differ by an amount of order $1/L$. We should therefore set $\delta$ to
to be roughly of this size. If we make $\delta>>1/L$,
even very uncompetitive species will be allocated some resources.
Hence the number 
of predator species per prey species will be large, and this will lead to a
highly connected web with a large number of links per species.
As $\delta \rightarrow 0$, only the main predator will be allocated
resources, and so in the limit the web will become a single food chain.
For most of the results given in this paper 
$\delta$ is in the range 0.05 - 0.2, which is comparable with $1/L = 0.1$.

It is useful to mention several features which have deliberately been 
excluded from the model. Firstly, there is no variation between individuals
of a given species, and there is no genetics. Species are simply represented 
by a list of phenotypic features, which represent average properties for all
members of that species. A model which included both many species and many
individuals per species, each with its own genotype and/or phenotype would
require enormous computational resources.
Secondly, even though speciation is an essential
part of the model, we do not attempt to 
consider the mechanism by which speciation
occurs. We simply suppose that species have an inherent tendency to diversify.
Speciation involves the establishment of reproductive isolation by some
means or another, and we cannot deal with this in the absence of a genetic
description of the species.
When a new species is created it differs by only one feature from the parent
species. However, this is intended to
represent a major change to the phenotype, which would
probably involve changes in more than one gene sequence, and possibly some
considerable alteration to the developmental biology of the organism. Thus
changing a single feature does not represent a single mutation, but is the
result of many changes at the genetic level. In addition
we make no distinction between sexual and asexual methods of reproduction,
and the model is intended to apply equally well to either case.

\section{Results - Structure of Model Food Webs}

The figures presented in the tables of results represent averages taken over
many webs. For each set of parameters several 
independent simulation runs were performed using different randomly generated
$m$ matrices and different random feature sets for the environment (species 0).
Each run was for 250,000 timesteps, with the exception of the runs
with $R = 10^{10}$, which were for 500,000 timesteps.
In each case, average quantities were measured during the second half the run.
The different runs for each parameter set were then averaged. 
Fluctuations in the number of species and the number of links
between different runs with the same parameter values 
were surprisingly large, with standard deviations being up to 50\% of the mean. 
Since these quantities increase in relation to one another, the
fluctuations in the number of links per species are much smaller (typically 
$\pm$ 0.2). 
Standard deviations in the percentages of $B$ and $I$ species are 
typically $7\% - 10\%$. We have omitted error estimates from the tables
for clarity. We are principally interested in 
the main trends in the web properties as the parameters are 
varied, rather than in precise values. We have checked that these trends
are significant. The quoted figures in the tables
apply to webs of trophic species. 

\begin{table}
\begin{centering}
\begin{tabular}{|c|c|c|c|c|}
\hline
\multicolumn{5}{|c|} {$R=10^3$} \\
\hline
& $\delta=0.05$ & $\delta=0.1$ & $\delta=0.15$ & $\delta=0.2$ \\
\hline
no. species & $40$ & $80$ & $290$ & $320$ \\
\hline
no. links &$55$& $150$& $800$ &$1250$ \\ 
\hline
links per species & $1.4$&$1.9$ & $2.7$&$3.8$ \\ 
\hline
average level & $1.7$ & $1.4$ & $1.2$ & $1.1$ \\ 
\hline
max level & $3.0$ & $3.0$ & $3.0$ & $3.0$ \\ 
\hline
B species $(\%)$ & $40$ & $56$ & $74$ & $78$ \\ 
\hline
I species $(\%)$& $59$&$44$ & $26$ & $22$ \\ 
\hline
T species $(\%)$& $1$ & $0$ & $0$ & $0$ \\ 
\hline
IB links    $(\%)$& $53$ & $73$ & $83$ & $88$ \\ 
\hline
II links    $(\%)$& $43$ & $27$ & $17$ & $12$ \\ 
\hline
TI links    $(\%)$& $4$ & $0$ & $0$ & $0$ \\ 
\hline
prey/predators & $1.6$ & $2.2$ & $3.8$ & 
$4.6$ \\ 
\hline
\end{tabular}

\caption{Results of a simulation of the model with
$\lambda=0.1$ and $R=10^{3}$ for four values of the competition
parameter $\delta$.}
\label{tab1}
\end{centering}
\end{table}

\begin{table}
\begin{centering}
\begin{tabular}{|c|c|c|c|c|c|}
\hline
& \multicolumn{4}{|c|} {$R=10^6$} & $R=10^{10}$ \\
\hline
& $\delta=0.05$ & $\delta=0.1$ & $\delta=0.15$ & $\delta=0.2$ & $\delta=0.05$ \\
\hline
no. species  & $150$ & $200$ & $410$ & $810$ & 220 \\
\hline
no. links  &$240$& $510$& $1420$ &$5930$ & 350 \\ 
\hline
links per species & $1.6$ & $2.5$ & $3.5$ & $7.3$ & 1.6 \\ 
\hline
average level & $2.5$ & $2.4$ & $2.3$ & $1.8$ & 3.2 \\ 
\hline
max level & $5.2$ & $5.2$ & $5.3$ & $5.5$ & 7.2 \\ 
\hline
B species $(\%)$& $10$ & $25$ & $32$ & $60$ & 3 \\ 
\hline
I species $(\%)$& $90$ & $75$ & $68$ & $40$ & 97 \\ 
\hline
T species $(\%)$& $0$ & $0$ & $0$ & $0$ & 0 \\ 
\hline
IB links    $(\%)$& $22$ & $47$ & $51$ & $70$ & 10 \\ 
\hline
II links    $(\%)$& $78$ & $53$ & $49$ & $30$ & 90 \\ 
\hline
TI links    $(\%)$& $0$ & $0$ & $0$ & $0$ & 0 \\ 
\hline
prey/predators & $1.1$& $1.3$&$1.5$ & $2.5$ & 1.0 \\ 
\hline
\end{tabular}

\caption{Results of a simulation of the model with
$\lambda=0.1$, $R=10^{6}$ or $10^{10}$ and various values of $\delta$.}
\label{tab2}
\end{centering}
\end{table}

Tables 1 and 2 show results for $R = 10^3$, $10^6$ and $10^{10}$,
 for $\delta$ in the
range 0.05-0.2. In all cases $\lambda = 0.1$. There is a clear tendency for 
the mean number of species to increase with $R$, for fixed values of 
$\delta$ and $\lambda$. However, it should be noted that the number of species 
is very much less than the theoretical maximum of $R$. In fact the number
of species only 
increases very slowly with $R$: 
a change in $R$ by seven orders of magnitude only 
causes a fivefold increase in the number of species. In contrast, the
number of species increases rapidly
 with $\delta$ 
at fixed $R$. Also, changing $\delta$ has a large effect on the total number 
of links. The number of links increases more rapidly than the number of 
species, so that the number of links per species increases significantly with 
$\delta$. The number of links per species increases slightly with $R$ for fixed 
$\delta$.

These results imply that competition is a significant factor in determining the
number of species and the number of links in the web. The competition parameter
$\delta$ determines the cut-off in the function $F_{n,n'}$ in equation
(\ref{compet}), and hence controls the number of predator species which can
successfully obtain resources from each prey. It is therefore to be expected
that the number of links per species will increase with $\delta$. Also, when 
the competition is weaker, it is easier for a species to find at least one
prey for which it is sufficiently well adapted to be a predator, therefore the
total number of species should also increase with $\delta$, as is observed. The
importance of $\delta$ in controlling species numbers also provides an 
explanation
of why the number of species only increases very slowly with $R$. The number of
level 1 species is limited by competition for external resources. Increasing $R$
at fixed $\delta$ tends to increase the populations of all level 1 species 
proportionately, rather than increasing the number of level 1 species. The 
same is also true of intermediate levels. However at the higher trophic levels 
species numbers are limited by the criterion of the minimum population size 
necessary for viability ($N(n)>1$). The principal effect of increasing $R$ is 
therefore to increase the maximum number of trophic levels possible in the web,
and to allow
a few high level species to survive, without changing the number of low level 
species very much. We have already shown that the maximum number of levels
increases logarithmically with $R$, and this therefore suggests that the total 
number of species should increase as $\log R$, which is consistent with our
results.
Figure 2 shows a histogram of the mean number of species at each level, and 
the way that this depends on $R$. This confirms the fact that increasing $R$ 
leads principally to an increase in species at higher trophic levels. 

Tables 1 and 2 
also show the change in the number of levels in the web with $R$ and 
$\delta$. The average level is defined by
calculating the level of each species, averaging these for all
species in each web, and then averaging these web averages over many webs for
each parameter set. The maximum level is defined by calculating the maximum
level present in each web, and averaging these maximum values over many webs for
each parameter set. Both average and maximum levels increase approximately 
logarithmically with $R$ as expected. For $R = 10^3, 10^6$ and $10^{10}$ the
maximum possible number of levels from equation (\ref{maxlevel}) are 3, 6 and 10.
For $R = 10^3$ the limiting value is always achieved. For $R = 10^6$ the mean 
value of the maximum level is only slightly below the limit, whereas for 
$R = 10^{10}$ it is considerably below. Two factors contribute to this.
Firstly, the theoretical limit was based on a
single food chain since this is the most efficient way of transferring resources
to high levels. The webs generated by the model are multiply connected and
contain omnivorous links and loops, hence the maximum attainable number of
levels is reduced. Secondly, the calculated limit takes only ecological 
efficiency into account, but ignores evolutionary constraints, which may also
limit the number of levels in the Webworld model. Well adapted level 1 species
may begin to evolve from the start of a simulation run, since the 
environment is fixed. High level species depend on lower level ones for their
resources, hence well adapted high level species tend to evolve at later stages
of the simulation when the properties of the lower level species are changing
less rapidly. The properties of the runs with $R = 10^{10}$ still 
appeared to be changing 
after 250,000 timesteps, hence they were continued for 500,000 
timesteps. This led to a significant increase in the number of species and the 
number of links, and a slight increase in the number of levels. Most of the 
late evolving species are on higher levels, hence the proportion of basal
species was found to decrease significantly between 250,000 and 500,000 
timesteps.

Also from the tables it can be seen that the average level
 decreases with increasing
$\delta$. This is understandable, because the level is defined as the shortest
possible chain to the external environment node. If $\delta$ is large, then a
species with a low score $S_{n,0}$, but a high score $S_{n,n'}$ against
another species $n'$ may successfully compete for the external resources,
and will therefore count as level 1 (the link to $n'$ would not affect the
calculation of the level). If $\delta$ is smaller, the same species
would only survive by being a predator of $n'$, and would therefore count as
a higher level species. The maximum number of levels does not change much
with $\delta$. The apparent slight increase in the maximum level with 
$\delta$ in 
table 2 is probably not significant.

The proportions of basal, intermediate and top species are given in tables 
1 and 2. Increasing $\delta$ at fixed $R$ leads to an increase in $B$ and a 
decrease in $I$ and $T$. This is for the same reason that the average level 
decreases with $\delta$. (However, in the following section we consider a
case with extremely large $\delta$ where this trend is no longer true).
Increasing $R$ at fixed $\delta$ leads to a decrease in $B$ and an increase in 
$I$. Again this is because increasing $R$ increases the number of levels, and 
thus the proportion of species in level 1 goes down. The number of top species 
is always very small: $T\leq 1.5\%$ when $R = 10^3$, and no top species were 
observed at all for $R = 10^6$ and $R = 10^{10}$. Thus increasing the number 
of levels does not mean that the number of top species
increases. In the model webs, for the parameters discussed so far, almost all
species have predator species, even if they are at high trophic levels. This
implies the presence of large numbers of loops in the food web and large 
numbers of omnivores. The behaviour of
$T$ is an important property which has been remarked upon in the study of real 
food webs, and we will return to it in the following section.

The tables also give the fractions of links between top, basal, and 
intermediate 
species, and the ratio of prey species to predator species. These quantities are
given primarily for comparison with the real food web data. Trends in these
quantities can be understood in terms of the trends discussed above in $B$, $I$,
and $T$. For example, as $\delta$ increases at fixed $R$, or $R$ decreases at 
fixed $\delta$, the percentage of species which are intermediate decreases and
as a consequence the fraction of the links which are between intermediate
species also decreases.

\section{Comparison with Real Food Webs}

The problems in getting reliable data on real food webs are readily
acknowledged by ecologists (Cohen et al. 1993). 
These problems centre around the 
tremendous amount of effort required to observe all species and all 
predator-prey interactions in a given ecosystem. Data from a large
number of food web studies has been assembled in the EcoWeb database 
(Cohen, 1989). Our analysis of the average properties of all the 181 webs 
in this database is given in Table 3. 
In addition we give data from several well studied webs which 
have been published recently (see caption to Table 3).

\begin{table}
\begin{centering}
\begin{tabular}{|c|c|c|c|c|c|c|}
\hline
\multicolumn{7}{|c|} {Experimental Data} \\
\hline
& ECOWEB & Lovinkhoeve & Coachella & St. Martin & Ythan & Little Rock \\
\hline
no. species & $16$ & $15$ & $29$ & $42$ & 83 & $93$ \\
\hline
no. links &$33$& $30$ & $262$ & $203$ & 398 & $1033$ \\ 
\hline
links per species & $2.0$& $2.0$ & $9.0$ & $4.8$ & 4.8 & $11.1$ \\ 
\hline
average level & $2.1$ & $2.5$ & $2.0$ & $2.1$ & 2.5 & $2.2$ \\ 
\hline
max level & $3.2$ & $3$ & $3$ & $4$ & 4 & $3$ \\ 
\hline
B species $(\%)$ & $21$ & $13$ & $10$ & $14$ & 5 & $13$ \\ 
\hline
I species $(\%)$& $49$ & $74$ & $90$ & $69$ & 59 & $86$ \\ 
\hline
T species $(\%)$& $30$ & $13$ & $0$ & $17$ & 36 & $1$ \\ 
\hline
TB links    $(\%)$& $10$ & $3$ & $0$ & $3$ & 1 & $0$ \\ 
\hline
IB links    $(\%)$& $29$ & $10$ & $13$ & $19$ & 10 & $9$ \\ 
\hline
II links    $(\%)$& $29$ & $57$ & $87$ & $53$ & 51 & $91$ \\ 
\hline
TI links    $(\%)$& $32$ & $30$ & $0$ & $25$ & 38 & $0$ \\ 
\hline
prey/predators & $0.89$ & $1.0$ & $1.11$ & $0.97$ & 0.67 & $1.13$ \\ 
\hline
\end{tabular}

\caption{Data from experimentally studied food webs: the ECOWEB database 
- average properties of all webs, (Cohen, 1989); 
the Lovinkhoeve Experimental Farm soil food web (De Ruiter et al, 1995);
the Coachella Valley desert (Polis, 1991); 
St. Martin Island (Goldwasser \& Roughgarden, 1993); 
the Ythan Estuary (Hall \& Raffaelli, 1991; Huxham et al., 1996);
and Little Rock Lake (Martinez, 1991). }
\label{tab3}
\end{centering}
\end{table}

Before these figures can be compared with the model it is necessary to note 
several points. For both the model and the real data, 
all statistics are for trophic species. Many webs contain detritus as a
`species', and we have followed the convention of treating this as a single
basal species. Plants are often treated in a very aggregated way, e.g. the
St. Martin Web divides plants into fruit, nectar, leaves, wood and roots, each
of which is treated as a `species'. Several webs split taxa into adult and 
juvenile `species' where these have different diets. 
Again we have followed the conventions of
the source article in these cases. 
In the model webs the external resources are treated
explicitly as a node in the web, and basal species must be linked to this node. 
Real webs have no such external resources node (although evidently external
resources do enter the real ecosystems). Therefore, when counting the number
of links in the model webs, links to the external resources node were omitted,
in order to allow comparison with the real webs. The real webs contain
some cannibalistic links (particularly the Coachella web). For simplicity we 
included these in the count of links, whereas cannibalism does not occur in the
model webs. The statistics in the table differ very little according to whether
one counts or discounts cannibalistic links. 

Most of the webs in the EcoWeb compilation are small, and they have been criticised
as being incomplete, and biased due to observational methods (Martinez, 1991;
Polis, 1991).
We have chosen the individual webs since in these cases attempts
have been made to be systematic and inclusive. Nevertheless, one suspects that
the large difference in the number of trophic species between these data sets
reflects the degree of aggregation chosen by the different workers, rather than
the actual diversity of the ecosystems. The five individual webs
are listed in order of increasing number of trophic species. There are some
trends apparent when comparing these webs with each other, although in many
ways each must be considered as a special case. 
There is a general trend for the number of links per species to be higher in 
larger webs. This would suggest that systematic study of a community over a 
long period of time leads to a greater increase in the number of links
observed than in the number of species. It may be that the number of links is
severely underestimated in some of the Ecoweb food webs. The Coachella web
breaks this trend, since it has a very large number of links per species and is
only fairly small. 
	
There is a strong similarity between the figures for the
Coachella and Little Rock webs, despite the difference in the number of species.
They both have a high number of links per species, a low value of $B$, a high
value of $I$ and very low $T$.
Martinez \& Lawton (1995) have suggested that, in the limit of extremely
large webs which represent large geographical areas, $I$ will increase to about
95\%, $B$ will decrease to about 5\%, and $T$ will decrease to zero. Our data in
tables 1 and 2 confirm that as $R$ increases, the number of species 
increases and the fraction of intermediates becomes larger ($I$ = 97\% in the 
run with $R = 10^{10}$, for which the average number of trophic species is
200). The data from the Ythan web appear to go against this trend, however, 
since there is a very high value of $T$ (36\%), even though the web is almost
as large as the Little Rock web. We have used the version of the Ythan 
web without parasites (Huxham et al. 1996). Many of the top species become
intermediate species when parasites are included, however the parasites 
themselves form a new top level which are not themselves preyed on. The 
number of top species is still very large in the Ythan web when parasites are
included, hence this cannot account for the large difference between the Ythan
and Little Rock webs.

When comparing the model results with the real data we view $R$ as a 
parameter which could change between ecosystems in different places, whereas
$\delta$ is a fundamental
property of the competition interactions between species 
which should presumably not vary much between locations. Therefore we would
wish to choose a value of $\delta$ which best fits a range of real data. 
The model results for $\delta = 0.05$ have high $I$, low $B$ and very low $T$,
which seems to be characteristic of the well-characterised real webs. All
the real webs have a maximum level of 3 or 4. We obtain this number of levels
in the model if $R = 10^3$ or $10^4$. The very high value of 7 for the maximum
level in the runs with $R = 10^{10}$ does not appear to be consistent with the
data, however it should be noted that all the model data is for $\lambda = 0.1$.
If we reduce $\lambda$ then the number of levels in the model will reduce, and
we will require larger values of $R$ in order to obtain 3 or 4 trophic levels.
Low values of $\delta$ seem to be preferred for matching the values of $B$ and
$I$ in the real data, however in this case the model predicts less than 2 links
per species. The number of links per species in the real webs tend to be higher
(as much as 11 in Little Rock). It is possible to get high numbers of links
per species in the model by increasing $\delta$ (see e.g. $R = 10^6$,
$\delta = 0.2$). However in this case we get much higher values of $B$ than
are seen in real webs. We conclude that the properties of the model webs are
generally of similar orders of magnitude to those of the real webs, but that
we have a not found a single set of parameters $R, \lambda$ and $\delta$ which
accurately match all the properties of the real data simultaneously. 
We have certainly not done an exhaustive search of parameter space yet for the
model: we did not change $\lambda$ at all. So it is possible that the 
agreement between the model and the data could be improved. 
Also, there are many refinements which we could make to the
model, such as changing the rules by 
which we distribute resources between species, or changing the way the scores
are calculated. 
Consideration of all these possibilities
would be justified if there were an excellent set of data on real food web 
structure against which to test the models. However, there are still problems
with even the best experimental data, and in view of this, it does not seem
worthwhile worrying too much about the precise values of the food web
statistics from the model. The model may in fact make some useful predictions
which are difficult to test in real life. For example, we noted above that
if $\delta$ is increased, so as to give high numbers of links per species, the
value of $B$ in the model appeared too high. Real webs tend to apply much
higher degree of aggregation to low level species than high level ones, so
maybe the values of $B$ in the real webs should really be higher than they are.

Over the range of $\delta$ of 
order $1/L$ considered in the tables, $I$ is a decreasing
function of $\delta$ and $B$ is an increasing function. However, as a limiting
case,
we also carried out a simulation with $R = 100$ and
$\delta = 25$. With this very large $\delta$ there were almost no extinctions 
due to competition. After 500 speciations there were approximately 90
different species which had average population sizes just greater than 1.
Extinctions occurred only due to the rule that a 
minimum population size of 1 is necessary for viability. In this run we
found that $I$ = 99\%
and $B$ = 1\%. This implies that there is a reversal of the trend in $B$ and $I$
for very large values of $\delta$. We do not think that this very large $\delta$
value is a reasonable one for comparison with real webs, however this simulation
does illustrate a limiting case of our model: in absence of competition very 
large highly connected webs arise which are almost entirely intermediate 
species.

We wish to comment further on the meaning of trophic levels in the Webworld 
model. The simple picture of ecosystems which is often envisaged supposes that
organisms can be grouped uniquely into trophic levels representing plants,
herbivores, carnivores, secondary carnivores etc. This picture supposes that
each level only interacts with the levels immediately above and below it.
Although it was argued from some early food web data that omnivores (i.e.
species feeding on more than one trophic level) were rare (Cohen et al., 1990),
it has since been argued that omnivory in well-characterized webs is frequent,
that organisms apparently assigned to the same level are far from
equivalent in their dynamics and interactions, and hence
that the concept of trophic levels is of little use (Polis \& Strong, 1996).
We found that omnivory is frequent in the model, and that it is not
possible to assign species to levels such that interactions are only with the
levels immediately above and below. However, we maintain that the idea of level
which we use here is a useful one, since it is a measure of food chain length:
the level of a species is the length of the shortest chain from the external
resources to the species (including the link to the external node).
 This is a different definition of chain length and
trophic level from that used in many other studies. For example, Martinez (1991)
calculates the average length of all food chains leading to each
species in the Little Rock web
using two algorithms which differ in the way they deal with
loops (it is necessary to exclude loops in some way or else 
chain length is infinite). He then defines the trophic level as the closest
integer to the average chain length + 1. Since there are huge numbers of chains
in large webs, algorithms which involve averaging all possible chains are slow.
Martinez was limited by computer time
 to looking at aggregated versions of his web using his
algorithm.  In analysing the model data we have thousands of webs, each with
hundreds of species, rather than just one real web, therefore we must use a
rapid algorithm for practical reasons. The maximum level is 3 in the Little
Rock trophic species 
web using our definition, whereas using Martinez's algorithm the maximum
level is 9. It is clear that the Little Rock web is extremely complex and that
long chains occur frequently, however, this is a statement about the 
combinatorial properties of highly connected graphs, and may not be
significant from an ecological point of view. Our algorithm shows that there is
no species in the Little Rock web 
which does not have a chain of length 3 or shorter. We believe that
shorter chains are much more important energetically than longer ones. This is
certainly true in the model webs, since there is an efficiency factor of
$\lambda = 0.1$ associated with every additional link. It is probably true
also in real webs, and could be tested using data which measures energy flow 
along each link. Therefore we believe that our algorithm for chain length and
trophic levels is justified ecologically as well as on grounds of practicality.

\begin{table}
\begin{centering}
\begin{tabular}{|c|c|c|c|c|}
\hline
\multicolumn{5}{|c|} {$\delta=0.05$, $R=10^{6}$} \\
\hline
& all links & $10 \%$ removed & $30 \%$ removed & $50 \%$ removed \\
\hline
no. species &$150$& $150 $& $140$& $130$   \\ 
\hline
no. links &$240$& $220 $& $170$& $125$  \\ 
\hline
links per species& $1.6$ & $1.5$ & $1.2$ &$1.0$ \\ 
\hline
average level & $2.5$ & $2.3$ & $2.3$ & $2.1$ \\ 
\hline
max level & $5.2$ & $4.5$ & $4.2$ & $4.0$ \\ 
\hline
B species $(\%)$ & $10$ & $16$ & $24$ &$30$ \\ 
\hline
I species $(\%)$& $90$ & $79$ & $57$ & $43$ \\ 
\hline
T species $(\%)$& $0$ & $5$ & $19$ & $25$ \\ 
\hline
TB links    $(\%)$& $0$ & $2$ & $8$ & $16$ \\ 
\hline
IB links    $(\%)$& $22$ & $25$ & $29$ & $31$ \\ 
\hline
II links    $(\%)$& $78$ & $68 $ & $47$ & $31$ \\ 
\hline
TI links    $(\%)$& $0$ & $5$ & $15$ & $21$ \\ 
\hline
prey/predators & $1.1$ & $1.2$ & $1.1$ & 
$1.0$ \\ 
\hline
\end{tabular}

\caption{The 
effect on the web properties of removing a fraction of the links 
at random.}
\label{tab4}
\end{centering}
\end{table}

Since real webs are likely to be incomplete due to observational difficulties,
Goldwasser \& Roughgarden (1997)
investigated the effect of incompleteness of data by
deliberately omitting
links from the original St. Martin island web. 
They have shown that most web properties vary
strongly as links are removed.
We carried out the following 
procedure in order to investigate the effect that incompleteness of data would 
have on the food webs in our model. Beginning with a set of webs generated 
from the model with $R = 10^6$ and $\delta = 0.05$, we deleted a certain 
proportion of links randomly. Deleted links represent real predator-prey 
interactions which were not observed due to poor experimental resolution.
After removal of the links,
if there were any species which were left unconnected to the rest of the web 
these species were also removed from the web. The deletion of the links was
carried out on the original species web, and the trophic species web for the 
new set of original species interactions was then found. 
As with all the tables in this article, figures refer
to the trophic species webs.
Table 4 shows the results of this process. $T$ increases rapidly as links are
removed, $I$ decreases rapidly, and $B$ increases more slowly. There are
consequent changes in the proportions of $TB$, $IB$, $II$ and $TI$ links. The
number of links per species, the average level and the maximum level all
decrease as links are removed. All these trends are the same as those observed 
when links are removed from the St. Martin web (Goldwasser \& Roughgarden,
1997).
 The proportions of $B$, $I$ and $T$ species after removal of a substantial
fraction of links
are similar to those in the EcoWeb data, whereas before removal of the links
they are close to the values of the more highly resolved webs like Little Rock. 
This supports the hypothesis that many of the webs in the EcoWeb collection 
suffer substantially from incompleteness. Real webs are also influenced by the
degree to which species are aggregated into clusters. Martinez (1991)
has shown that aggregation of species within the original Little Rock web 
results in smaller webs with statistics which are closer to those of the
EcoWeb compilation than the full web, which suggests that the smaller webs
are also influenced significantly by aggregation. We have not yet considered 
aggregation effects in our model webs, but we expect that the changes will be 
similar to those observed with the Little Rock web. 

\section{Results - Evolutionary Dynamics}

Figure 3 shows the number of species as a function of
evolutionary time for three runs of the simulation with the same parameter
values. The numbers of species differ significantly between different
random assignments of the $m_{ij}$ matrix.
In each case the number of 
species tends to increase fairly rapidly at first, but after a certain time 
the system tends to a steady state with fairly constant species number. 
Extinction avalanches are visible as sudden drops in the species number. The 
figure shows the number of original species, but the curves for the number of
trophic species follow the same pattern with slightly smaller numbers, and show
extinction events occurring in the same places. In
previous models of macroevolution, attention has focused on the dynamics of 
extinction events when the model is in a stable state. The avalanche size 
distribution has been measured and has been observed to have a power law shape 
for some of these models (Bak \& Sneppen, 1993;
Sol\'e \& Bascompte, 1996). In the Webworld model
we may define the avalanche size at a given timestep as the number of
species which become extinct due to the addition of one new species (if the new
species adds stably to the web this is an avalanche of size zero). We intend to 
discuss the sizes of extinction events in our model in more detail in a 
subsequent paper. However, our initial results suggest that the size of 
avalanches tends to decrease with time in the Webworld model.
Once the system has reached a steady state, fluctuations in 
species number tend to be small, whilst in 
the initial stages of the simulation 
large avalanches can occur. Within the model large avalanches are associated with
evolutionary progress. Initially there are relatively few species, and these are
relatively poorly adapted. If we consider only level 1 species initially, 
then we expect that the scores $S_{n,0}$ of species will gradually increase
as better adapted primary producers evolve. In the early stages of the
simulation there is a reasonable chance that a new species will be substantially
better than existing ones, and an extinction avalanche may occur when the new
species evolves. As time goes on it will become
increasingly more difficult to evolve species which are better adapted, and
when improvements do occur, scores are likely to increase by smaller amounts. 
Hence avalanches are likely to decrease in size and in frequency as time goes 
on. Since the level 1 species tend to change less rapidly as they become 
better adapted it follows that conditions for higher level species become more 
stable. Hence level 2 species can become increasingly better predators of the 
existing level 1
species, and so on throughout the web. Therefore evolutionary changes on all 
levels of the web are likely to slow down in the same way. 

Self-organized models of evolution are designed in such a way that the 
stationary state is critical and has interesting dynamics. In these models the 
avalanche size distribution and the species lifetime distribution are power 
laws. We believe that after a very long time our model would 
reach an absolutely stationary state where it would be impossible for any new 
species to evolve. None of our simulations ever reached this point entirely, 
although the probability that a new species survives on the first timestep
in which it evolved became very small toward the end of the runs. 
The time to reach this point 
of evolutionary `stagnation' should depend on the total number of possible 
species in the model. The number of combinations of 10 features out of 500 
possible ones is approximately $2.5 \times 10^{20}$, which is large but still 
finite. We intend to investigate the effect of changing the total number
of possible features $K$ in future work.

The initial period of evolutionary progress and large avalanche sizes is an 
interesting feature of our model, and we believe it may tell us something 
about the real world. Is the real world in a stationary state? Are the 
properties of ecosystems today statistically equivalent to those in earlier 
geological periods? We do not know how to answer these questions, but we 
believe they are important questions to ask. It is clear that the model has a 
finite number of possible species, but it is not clear whether this is true 
for the real world. The earth is obviously much more complex than any computer 
model, however it is still finite in terms of the available space, energy and 
raw materials. If the earth were unchanging, it does not seem unreasonable to 
suppose that real ecosystems would gradually perfect themselves to the 
external conditions, and that the consequent rate of evolutionary change would 
decrease, just as in the model. The origin of life probably occurred about 
$3.5-3.8\times 10^9$ years ago. It is not clear whether this is a long or a 
short period measured on the timescale of what is evolutionarily possible. 
Measures of diversity in the fossil record (Sepkoski, 1993)
show a general increasing trend in species numbers between the Cambrian and the
present, despite several very large extinction events, and there is no real
indication of any levelling off in diversity.
 
Of course, the conditions on earth are not fixed: climatic change occurs on a 
wide range of timescales. If we view the non-living world as continually 
changing, then the living world must also continually change, and never has the 
opportunity to reach evolutionary stagnation. There is also the possibility 
that the external conditions might change by sudden rare cataclysmic events, 
such as meteorite strikes, rather than by smooth gradual change. Such a large 
scale external event might cause a large scale extinction, and might change 
conditions sufficiently that the evolutionary clock would effectively be set 
back to an early stage where ecosystems were poorly adapted to the environment.
This would enable a new burst of evolution, with many novel species arising. 
This suggests a picture of the real
world where there is a considerable rate of inherent evolutionary change and 
considerable fluctuation in species numbers, and where external events causing
major changes happen
sufficiently often to prevent evolutionary stagnation from 
occurring. The data in the fossil record may some-day be complete enough to 
give some answers these questions. In the meantime there are still many 
useful issues which can be addressed by studying models such as ours.

\section{Conclusions}

The Webworld model describes the interactions of
coevolving species. The model makes predictions concerning the structure of 
food webs which can be compared with data on real webs. There are only 
three parameters in the model --- $R$, $\lambda$, and $\delta$ --- hence, the 
number of testable predictions is much larger than the number of parameters. 
The measured quantities such as the number of links per species, the number of 
trophic levels,
and the proportions of top, basal and intermediate species, are not far from the
values observed in real food webs in most cases. It would be possible to choose 
parameters so that the results match a particular set of real food web data as
closely as possible. 
However, given the uncertainties in most of 
the present food web data, we have not tried to match these data too closely. 
Instead we have tried to point out the major qualitative trends in food web 
properties which occur when the parameters are changed. These
trends make sense from an ecological point of view. 
We would like to contrast our model with the
cascade model of food web structure (Cohen, 1990; Cohen et al., 1990). 
Even though the 
cascade model successfully describes many food web properties, it is
basically just a set of probabilistic rules for assigning links between nodes
in a graph. The justification of these rules comes entirely from comparison
model webs with real data. In contrast, the parameters in our model
have an ecological meaning: the available external resources, the fraction of
resources transferred from prey to predator, and the strength of competition
are all meaningful quantities in real webs.

An important question in food web theory which has had considerable attention
recently is the issue of food web assembly (Luh \& Pimm, 1993;
Morton \& Law, 1997). In assembly
models species are
invading the ecological community from an external species pool and are
therefore unrelated to existing species, whereas in the
Webworld model new species are arising by evolution and are therefore similar
to the species from which they evolve. We intend to develop the model to
compare properties of webs generated by invasion and by evolution. One
observation in assembly models is that the probability of successful 
invasion of a community tends to decrease with time.
This seems to have a parallel in Webworld, where 
the survival probability of a newly generated species tends to decrease with
time.

This work was motivated in part by the wish to explore recent claims that
models of evolution may have a tendency to ``self-organise" into a critical
non-equilibrium state which has avalanches on all scales. We approached the 
problem by trying to design a realistic model for co-evolution in which the
evolutionary dynamics can be studied, rather than by deliberately designing 
a very simple model with interesting dynamics, but which is more difficult to
relate to real evolutionary and ecological phenomena. This article has 
concentrated on the ecological properties of the food webs, and we are currently
investigating the evolutionary properties of Webworld in more detail.

\section*{Acknowledgements}

We thank Mark Huxham for clarifying some details
regarding the Ythan web.
This work was supported in part by a grant from the University of Manchester
and by EPSRC grant GR/K/79307. 

\section*{References}

\noindent Arditi R. \& Michalski, J. (1996) Nonlinear Food Web Models
and their Responses to Increased Basal Productivity. {\em Food Webs: Integration
of Patterns and Dynamics} pp 122-133. Eds. G. A. Polis and K. O. Winemiller.
Chapman and Hall, New York.

\medskip

\noindent Bak, P., Tang, C. \& Wiesenfeld, K. (1988) Self-organised
criticality. {\em Phys. Rev. A} {\bf 38}, 364-374.

\medskip

\noindent Bak, P. \& Sneppen, K. (1993) Punctuated equilibrium and
criticality in a simple model of evolution. {\em Phys. Rev. Lett.} {\bf 71}, 
4083-4086.

\medskip

\noindent Berryman, A.A., Michalski, J., Gutierrez, A.P. \& Arditi, R.
(1995) Logistic theory of Food Web Dynamics. {\em Ecology} {\bf 76}, 336-343.

\medskip

\noindent Briand, F. \& Cohen, J.E. (1984) Community food webs have a
scale-invariant structure. {\em Nature}, {\bf 307}, 264-267.

\medskip

\noindent Cohen, J.E. (1989) {\em Ecologists' Co-Operative Web Bank.
Version 1.00.  Machine-readable data base of food webs}.  New York:
The Rockefeller University.

\medskip

\noindent Cohen, J.E. (1990) A stochastic theory of community food webs.
VI. Heterogeneous alternatives to the cascade model.{\em Theor. Pop. Biol.} 
{\bf 37}, 55-90.

\medskip

\noindent Cohen, J.E., Briand, F. \& Newman, C.M. (1990)
{\em Biomathematics}, {\bf 20} \lq\lq Community Food Webs, Data and Theory", 
Springer Verlag, and references therein.

\medskip

\noindent Cohen, J.E., Beaver, R.A., Cousins, S.H., DeAngelis, D.L.,
Goldwasser, L., Heong, K.L., Holt, R.D., Kohn, A.J., Lawton, J.H., Martinez, N.,
O'Malley, R., Page, L.M., Patten, B.C., Pimm, S.L., Polis, G.A., Rejmanek, M.,
Schoener, T.W., Schoenly, K., Spules, W.G., Teal, J.M., Ulanowicz, R.E.,
Warren, P.H., Wilbur, H.M. \& Yodzis, P. (1993) Improving Food Webs.
{\em Ecology} {\bf 74}, 252-258.

\medskip

\noindent De Ruiter, P.C., Neutel, A.M. \& Moore, J.C. (1995) Energetics,
Patterns of Interaction Strengths, and Stability in Real Ecosystems. 
{\em Science} {\bf 269}, 1257-1260.

\medskip

\noindent Goldwasser, L. \& Roughgarden, J. (1993) Construction and 
analysis of a large Caribbean food web. {\em Ecology}, {\bf 74}, 1216-1233.

\medskip

\noindent Goldwasser, L. \& Roughgarden, J. (1997) Sampling effects and the
estimation of food web properties. {\em Ecology} {\bf 78}, 41-54.

\medskip

\noindent Hall, S.J. \& Raffaelli, D. (1991) Food-web patterns: lessons 
from a species-rich web. {\em J. Anim. Ecol.} {\bf 60}, 823-842.

\medskip

\noindent Huxham, M., Beaney, S. \& Raffaelli, D. (1996) Do parasites reduce
the chances of triangulation in a real food web? {\em Oikos} {\bf 76}, 284-300.

\medskip

\noindent Kramer, M., Van de Walle, N. \& Ausloos, M. (1996) Speciations
and Extinctions in a Self-Organizing Critical Model of Tree-like Evolution.
{\em J. Phys. I France} {\bf 6}, 599-606.

\medskip

\noindent Luh, H.K. \& Pimm, S.L. (1993) The assembly of ecological
communities: a minimalist approach. {\em J. Anim. Ecol.} {\bf 62}, 749-765.

\medskip

\noindent Martinez, N.D. (1991) Artifacts or attributes? Effects of 
resolution
on the Little Rock Lake food web. {\em Ecol. Mono.} {\bf 61}, 367-392.

\medskip

\noindent Martinez, N.D. \& Lawton, J.H. (1995) Scale and food web 
structure --- from local to global. {\em OIKOS} {\bf 73}, 148-154.

\medskip

\noindent Morin, P.J. \& Lawler, S.P. (1995) Food web architecture and
population dynamics: theory and empirical evidence. {\em Annu. Rev. Ecol. Syst.}
{\bf 26}, 505-529.

\medskip

\noindent Morton, R.D. \& Law, R. (1997) Regional Species Pools and the
Assembly of Ecological Communities. {\em J. Theor. Biol.} {\bf 187}, 321-331.

\medskip

\noindent Paczuski, M., Maslov, S. \& Bak, P. (1996) Avalanche dynamics in
evolution, growth and depinning models. {\em Phys. Rev. E} {\bf 53}, 414-443. 

\medskip

\noindent Pimm, S.L. (1982) {\em Food Webs} Chapman and Hall, London.

\medskip

\noindent Pimm, S.L., Lawton, J.H. \& Cohen, J.E. (1991) Food web 
patterns and their consequences. {\em Nature}, {\bf 350}, 669-674.

\medskip

\noindent Polis, G.A. (1991) Complex Trophic Interactions in Deserts: 
An Empirical Critique of Food Web Theory. {\em American Naturalist} {\bf 138},
123-155.

\medskip

\noindent Polis, G.A. \& Strong, D.R. (1996) Food Web Complexity
and Community Dynamics. {\em American Naturalist} {\bf 147}, 813-846.

\medskip

\noindent Roberts, B.W. \& Newman, M.E.J. (1996) A model of Evolution
and Extinction. {\em J. Theor. Biol.} {\bf 180}, 39-54.

\medskip

\noindent Sepkoski, J.J.Jr. (1993) Ten years in the library: new
data conform paleontological patterns. {\em Paleobiology} {\bf 19}, 43-51.

\medskip

\noindent Sneppen, K., Bak, P., Flyvbjerg, H. \& Jensen, M.H. (1995)
{\em Proc. Nat. Acad. Sci. USA} {\bf 92}, 5209-5213.

\medskip

\noindent Sol\'e, R.V. \& Bascompte, J. (1996) Are critical phenomena
relevant to large-scale evolution? {\em Proc. Roy. Soc. Lond.} 
{\bf B 263}, 161-168.

\medskip

\noindent Sol\'e, R.V., Bascompte, J. \& Manrubia, S.C. (1996) Extinction:
bad genes or weak chaos? {\em Proc. Roy. Soc. Lond.} {\bf B 263}, 1407-1413.

\medskip

\noindent Sol\'e, R.V., Manrubia, S.C., Benton, M. \& Bak, P. (1997)
Self-similarity of extinction statistics in the fossil record. {\em Nature}
{\bf 388}, 764-767.

\medskip

\noindent Vida, G., Szathmary, E., Nemeth, G., Hegedus, G., Juhasz-Nagy, P. \&
Molnar. I. (1990) Towards modelling 
community evolution: the Phylogenerator. {\em Organizational Constraints
on the Dynamics of Evolution.} pp 409-422. Eds. J. Maynard Smith and G. Vida.
Manchester University Press.

\medskip

\noindent Zheng, D.W., Bengtsson, J. \& Agren, G.I. (1997) Soil Food Webs
and Ecosystem Processes: Decomposition in Donor Control and Lotka-Volterra
Systems. {\em American Naturalist} {\bf 149}, 125-144.

\newpage

\begin{figure}
\centerline{\psfig{file=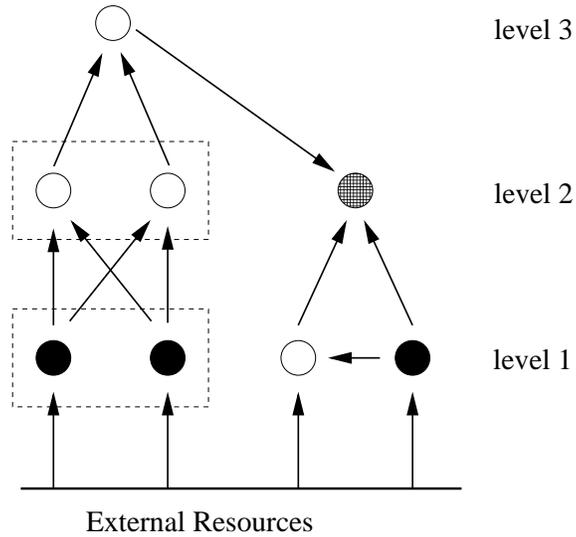,height=7.0cm}}
\label{fig1}
\caption{
An illustrative example of a food web. 
Arrows indicate the direction of flow of resources. Input
of external resources is specifically indicated. Basal species are black, 
intermediate species are white, and top species are patterned. Boxed species
form part of the same trophic species. The trophic level of a species is the
length of the shortest food chain from the external resources to that species.
}
\end{figure}

\begin{figure}
\centerline{\psfig{file=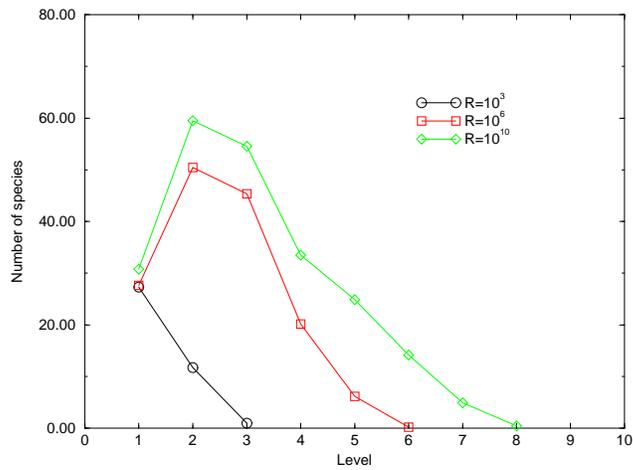,height=7.0cm,angle=270}}
\label{fig2}
\caption{
The mean number of species on each trophic level with $\lambda = 0.1$, 
$\delta = 0.05$ and $R = 10^3, 10^6$ and $10^{10}$.
}
\end{figure}

\begin{figure}
\centerline{\psfig{file=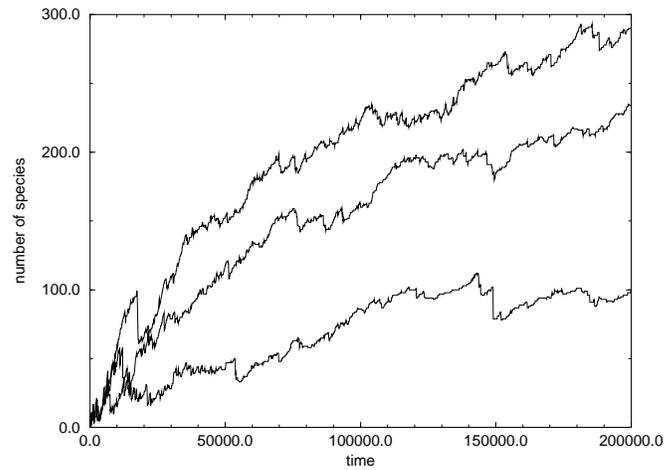,height=7.0cm,angle=270}}
\label{fig3}
\caption{
The number of species in the web is shown as a function of time for three
different runs of the Webworld model with the same parameters $R = 10^6$, 
$\delta = 0.05$ and $\lambda = 0.1$. }
\end{figure}

\end{document}